\documentclass[11pt]{article}
\usepackage{astrobib,epsf,aaspp4,psfig}
\def\lap{\lower.5ex\hbox{$\; \buildrel < \over \sim \;$}}
\def\gap{\lower.5ex\hbox{$\; \buildrel > \over \sim \;$}}
\begin{document}
\title{Comparing Young Stellar Populations Across Active Regions in the M31 Disk} 
\author{ Benjamin F. Williams}
\affil{University of Washington}
\affil{Astronomy Dept. Box 351580, Seattle, WA  98195-1580}
\affil{ben@astro.washington.edu}

\begin{abstract} 

I present preliminary BV resolved stellar photometry of the M31 disk
measured from the data set of the NOAO/CTIO Local Group Survey.  I
have performed detailed analyses of the star formation histories in
and around three currently active regions in the M31 disk: OB~78,
OB~102 and the northeast spiral arm.  The results suggest that low
H$\alpha$ emission from OB 78 compared to other active regions is
directly related to the details of the recent star formation histories
of the regions.  In addition, while every active region I analyzed
shows a recent increase in star formation activity, some active
regions also contain overdensities of stars with ages $\lap$100 Myr
compared to adjacent regions.  The adjacent regions show a relatively
constant, low star formation rate over the past 100 Myr.  There is no
significant difference between the stellar populations on either side
of the active regions.  This symmetry provides no obvious signature of
recent propagation of star formation near these currently active
regions of the M31 disk.

\end{abstract}
\keywords{galaxies: M31; spiral disks; stellar populations.}

\section{Introduction}

The stellar populations of the M31 disk provide an excellent
laboratory in which to measure the progression of star formation
through a large galactic disk.  Resolved photometric studies of the
stellar populations in the disk structures, in conjunction with
accurate models of stellar evolution, allow the star formation
histories (SFHs) of these structures to be measured.  Efforts to
disentangle the stellar populations in the M31 disk date back to
\citeN{hubble1929} who first resolved the stars, and
\citeN{baade1944}, who first noticed the multiple stellar populations.

Stellar photometry of the M31 disk has provided clues about the age
structure of its young stellar populations. \citeN{ivanov1985} used a
handful of the brightest stars in several of the southern OB
associations to suggest an age structure in the southern disk.
\citeN{haiman1994} found evidence that the populations of the OB
associations in the eastern and western spiral arms differ, but they
were unable to measure ages.  \citeN{humphreys1979} found the
characteristics of the brightest stars in the outer parts of M31
similar to those in our region of the Galaxy, suggesting similar
evolution scenarios.  \citeN{hodge1988} found the upper main sequence
luminosity function to be similar across the disk, indicating only
slight differences in recent SFHs. Furthermore, the age distribution
of the Cepheid variables and compact star clusters in the southern
disk of M31 are suggestive of a recent wave of star formation
propagating through the disk, currently located at NGC 206
(\citeNP{magnier1997a}, \citeNP{williams2001a}).  In addition,
infrared work by \citeN{kodaira1999} has provided evidence for
episodic star formation in different subregions of the OB association
A24 during each spiral wave passage \cite{kodaira1999}.

In the past few years several groups have used the Hubble Space
Telescope (HST) to put constraints on the metallicity and age
distributions of the stars in the M31 disk.  \citeN{magnier1997b}
found the age and reddening distributions of several OB
associations. \citeN{williams2001b} performed an automated search for
young open clusters in portions of these complexes, finding them to
have a wide range of reddenings and ages. \citeN{sarajedini2001} found
strong evidence for a thick disk population in an HST field in the M31
disk.  They found the populations similar to the Galaxy, although the
metallicities of the populations were higher.  \citeN{ferguson2001}
analyzed an HST field in the outer disk of M31, finding that it was
dominated by an old population with an intermediate
metallicity. \citeN{williams2002} studied 27 HST fields in the M31
disk taken from the HST archive, also seeing an intermediate age,
potential thick disk component in all fields.  In addition,
\citeN{williams2002} found that the disk is dominated by an older
population high in metallicity, confirming that the current star
formation rate of the disk is low compared with $\gap$1 Gyr ago.

While these recent HST studies have been essential for
understanding the stellar populations of the M31 disk, they were
unable to search for large scale star formation patterns in the disk.
While the older populations are well mixed and can be used to draw
general conclusions about the disk's early evolution from pointed
observations, the young population is still near the location of its
formation and is therefore much more structured within the disk.
Making sense of this structure in the young stellar populations
requires a data set with large angular coverage to probe for
propagation of star formation through the disk.  Such a study of the
young populations in the M31 disk is not easily done with the small
field of view of HST because each field only contains a small sample
of the youngest stars.  Until recently, such a study was not easily
done from the ground because of the limited field of view of most high
resolution ground-based imagers; however, with newly-developed,
multi-CCD, wide-field imagers available on large telescopes, it is now
possible to obtain high-resolution, homogeneous data sets covering
large fields.  This ability has allowed us to compare the stellar
populations in different regions of the disk with confidence that our
conclusions will not be due to systematic differences within the data
set.

In this paper, I use data from the recently conducted Local Group
Survey (LGS) \cite{massey2001} to analyze the stellar populations of
several key regions in the M31 disk: OB 102, OB 78
\cite{vandenbergh1964a}, and the northeastern spiral arm.  These
regions were chosen as ideal for understanding how star formation
progresses and propagates through a complex galactic disk.  The large,
homogeneous, high-resolution stellar photometry allows the
reconstruction of the recent SFHs across these active regions.
Results are compared with areas that have had their SFHs examined with
HST data in order to look for consistency.

\section{Data Acquisition and Reduction}

\subsection{Photometry}

All of the ground-based data used for this project were generously
supplied by the Local Group Survey (LGS) collaboration
(\citeNP{massey2001}; http://www.lowell.edu/users/massey/lgsurvey)
which is acquiring 1'' resolution, photometric data with the 8k x 8k
Mosaic cameras on the 4m telescopes at KPNO and CTIO, entirely
covering ten LG galaxies in U, B, V, R, I, and narrow-band H$\alpha$,
[S II] ($\lambda\lambda$6717,6731), and [O III] ($\lambda$5007).  The
LGS is working on providing their own, more rigorous, calibration of
these data, leading towards a complete UBVRI catalog of stars.
However for the purposes of the present paper I draw upon photometry
in the literature for calibration.  The modeling effort requires only
10\% photometry, which can readily be achieved by the means described
here.  These data will soon be, but are not yet, public.

I obtained the data for six Mosaic fields in the M31 disk from the LGS
team.  These data consisted of 5 dithered frames in B and V for each
field.  Each dithered frame had been flat-fielded, bias-corrected, and
the geometric distortions removed by the LGS team.  Fields F1, F2 and
F3 cover the northern disk; fields F8, F9 and F10 cover the southern
disk.  The B and V observations of each of the regions of interest
were completed in the same hour.  In order to measure the photometry
from the images efficiently, the dithered frames for each field were
aligned and stacked.  Because the B and V observations of each field
were completed within minutes of one another, photometry of variable
stars in the stacked images is similar to that of a single snapshot of
their changing colors and luminosities.  The stacked B band images of
the regions of interest for this study are marked and labeled in
Figure 1.

All of the photometry for this project, including PSF fitting,
measurement of instrumental magnitudes and execution of artificial
star tests, was done using the automated photometry routines DAOPHOT
II and ALLSTAR \cite{stetson}. Because the chips of the Mosaic camera
are not precisely coplanar, point-spread-function (PSF) fitting was
performed on each chip individually.  I tested the accuracy of the PSF
fitting photometry on the stacked images by running the same routine
on a single exposure and comparing the results.  Figure 2 shows the
results of this comparison for the B and V photometry.  The stacking
had only a negligible systematic effect on the measured magnitudes.
The mean was shifted by only 0.01 mag in B and 0.002 mag in V at the
bright end.

The zero point offsets for the photometry were determined
 statistically, matching stars from the catalogs of
 \citeN{magnier1992} and \citeN{mochejska2001}.  Field F2 overlapped
 with fields F1 and F3, and field F9 overlapped with fields F8 and
 F10.  In cases where there were few stars on which to base the B and
 V zero points (the outermost fields, F1 and F10, used for the
 foreground sample), I found duplicate measurements of stars in the
 overlapping field with a well-sampled zero point, and I set the zero
 point to match the photometry for these stars.  In these cases, I
 quote the root of the sum of the square error from the original zero
 point measurement and the square error from the dispersion from the
 separate measurements.  I also corrected each zero point by applying
 the color terms of the LGS\footnote{see
 http://www.lowell.edu/users/massey/lgsurvey/colorterms.html} to the
 mean B-V color of the calibration stars.  This correction factor
 accounts for the effect of the Mosaic color terms on the zero point
 measurement.  Table 1 provides the final zero points used and the
 standard deviation along with the number of stars used for the
 determination for each of the six fields.  Although the zero points
 have fairly large errors associated with them, the large number of
 stars that have gone into these zero points provide a robust mean.
 Outliers were removed using Chauvenet's criterion.  In these crowded
 fields, some fainter stars can be mistakenly matched to bright
 neighbors.  Chauvenet's criterion was an effective way of removing
 such mismatches.  Most of the dispersion about this mean is most
 likely due to the older catalogs, which often had less homogeneous
 data sets.

In order to test the accuracy of the zero points, I found stars
measured independently in fields F2 and F3, the fields with the
greatest difference in V-band zero point.  I corrected these stars
with the appropriate zero point and looked at the residuals.  These
residuals are plotted in Figure 3.  The figure shows that the zero
point offsets were effective.  The independent measurements had rms
scatter of 0.041 in B and 0.046 in V at the bright end.  Having
measured these statistics, I believe that this relative photometry
marks one of the most complete and homogeneous set of stellar
photometry in the M31 disk to date in B and V.

After obtaining the mean zero points for each field in each color, the
photometry of each star was corrected for the color terms of each
chip\footnote{see
http://www.lowell.edu/users/massey/lgsurvey/colorterms.html}.  These
corrections resulted in the final colors and magnitudes used for the
analysis.  While this photometry is preliminary to the final version
that will be measured by the LGS, the experiments described later
reveal that it is sufficiently precise for my purposes.  I dissected
these data in order to examine several key regions of the M31 disk.

In order to determine accurate SFHs from the photometry, it is
extremely important to have excellent statistics on the completeness
and measurement errors as a function of B-V color and V magnitude.  In
order to achieve this goal, I ran large numbers of artificial star
tests in each region under consideration.  Each test had to be run
separately because of the large differences in the effects of crowding
in different regions of the galaxy.  I found that I could obtain
accurate statistics from a set of 150,000 artificial stars in each
field section.  These tests were performed by adding a unique set of
1500 stars to each of a set of 100 copies of the image section in
question.  Each of these 100 copies were then run through my
photometry routine in order to determine completeness and errors.
Plots of random subsets of the artificial star results for the three
OB 102 field sections are shown in Figure 4.  The plots show only 5
percent of the results to make the figure easier to read; these
subsets are entirely random and representative of the full set of
tests.  The errors are mainly random Poisson errors; however, there
are some stars with large, negative residuals from crowding.  These
stars were added too close to a brighter star already in the image,
causing an over-estimation of the brightness.

\subsection{Foreground Contamination}

If confused for massive stars in the M31 disk, bright stars in the
 Galactic foreground could cause systematic errors in the measurements
 of recent SFHs.  These foreground stars were statistically subtracted
 from the CMDs of the regions under consideration.  A sample of
 foreground stars was taken from the far corners of the fields with
 the highest galactocentric distance (F1 and F10).  These foreground
 sampling areas covered 235 arcmin$^{2}$.  The combined CMD of these
 foreground-dominated areas is shown in Figure 5.  The main sequence
 of the M31 disk at B-V $\sim$0 is still visible, but weaker compared
 to the foreground sequence.

In order to subtract the proper number of foreground contaminants from
each analyzed region, I binned the CMD of the field sample.  The bin
sizes were 0.1 in B-V color and 0.15 in V magnitude.  I then scaled
the number of stars in each bin by the size of the area of interest
and subtracted that number of stars from its counterpart bin in the
data from the area of interest.  The number of foreground stars in
each region proved to be $\lap$3 percent of the stars in the region,
so that the details of the foreground sample had no significant impact
on the results of the analysis.

\section{Description of the Star Formation History Analysis Technique}

I determined the SFHs of the selected regions using the analysis
package MATCH (\citeNP{dolphin1997}, \citeNP{dolphin2000b}).  This
recently developed software uses a technique pioneered by earlier,
non-computationally intensive work on stellar populations in other
Local Group galaxies (e.g. NGC 6822, \citeN{gallart1996}) in which the
CMD of the stars was broken down into bins of color and magnitude.
These CMDs are also known as binned Hess Diagrams.  Before powerful
computing was possible, these bins were chosen to isolate populations
of differing ages and metallicities, and based on the relative number
of stars in each bin, a rough measurement of the star formation and
chemical enrichment histories of the field could be made.  MATCH takes
this concept to its full potential by using the most recent computing
power to create high-resolution Hess Diagrams of rectangular bins of
constant size specified by the user.

Once the CMD of the data has been transformed into a Hess diagram, the
software uses the stellar evolution models of \citeN{girardi2000} to
create a unique Hess diagram for each of a range of stellar ages and
metallicities.  These model-based diagrams are produced by populating
the CMD along the theoretical isochrone of each metallicity and age,
assuming a 1 $M_{\odot}/yr$ star formation rate and a Salpeter initial
mass function (IMF) and taking into account the completeness and
photometric errors in each bin of the CMD as measured from the
artificial star tests on the real data.  The program then finds the
linear combination of model-based Hess diagrams which best reproduces
the Hess diagram of the observed stars for each of a range of
reddening and distance values.  Since each model-based diagram was
created assuming a 1 $M_{\odot}/yr$ star formation rate, the
coefficients of this linear combination provide the SFH in units of
$M_{\odot}/yr$.  Finally, the best fits for each distance and
reddening value are weighted by goodness of fit to determine values
and errors for the most likely distance and extinction to the field as
well as values and errors for the star formation rates and metal
abundance spreads during each time period explored.  Finally, in order
to check the viability of the result, the package can be used to
create a synthetic CMD from the stellar evolution models by populating
the theoretical isochrones using the assumed IMF along with the
derived best fit for the metallicities and star formation rates for
each time period, taking into account the best fit distance and
reddening values for the field.  The artificially generated CMD can
then be compared to the observed CMD in order to verify that the
statistically determined SFH creates a stellar population similar to
the observed population.

I tested this technique for the situation of ground based data of the
M31 disk which suffers from differential reddening across and through
the disk as well as severe crowding.  Artificial star samples similar
to the data in star number and photometric accuracy, but with known
SFHs, were created.  Then the SFHs of these fake star fields were measured
and compared to the input SFHs.  I performed this experiment on two
different synthetic SFHs.  The results from the first are shown in
Figure 6.  This synthetic star sample was created with a low, constant
star formation rate until a very recent burst.  The metallicity was
kept constant at solar abundances, and the reddening was set to $A_V =
0.83$.  This reddening value was chosen as a random value in the range
the analysis had revealed to be typical ($0.3\lap A_V\lap 0.9$).
Figures 6a and 6b show the results of applying our analysis technique
directly to the synthetic sample.  The age distribution is
well-recovered, but the chemical evolution is not.  This inability of
the technique to find the right metallicity distribution has been seen
before (see \citeNP{williams2002}), and it is likely due to the
insensitivity of the color of the upper main sequence to metallicity.
Figures 6c and 6d show the results for the same artificial sample
after each star was assigned a random reddening value in the range
$1.23 > A_V > 0.43$.  Despite the scrambling of the colors and
magnitudes from differential reddening, the analysis technique
measures the right age distribution, revealing that the measurable
problems with the photometry from crowding are more important to take
into account than the effects of differential reddening.  The same
experiment was run for a SFH with two star forming bursts.  This age
distribution was also accurately recovered, as shown in Figures 7a -
7d.

\section{Results:  SFHs of Selected Regions}

Once convinced that meaningful age distributions could be retrieved
from the data, I obtained star samples from interesting areas of the
M31 disk to look for recent patterns of star formation around the most
active portions of the disk.  These areas of interest were chosen
because they stand out as large groupings of young stars likely
associated with the action of the spiral arms within the disk. The
results allow us to draw conclusions about the behavior of the spiral
arms in M31 over the past $\sim$200 Myr.  The distance moduli and mean
extinction corrections determined for each region analyzed are given
in Table 2.  The distance measurements are all the same within the
errors, which shows the consistency of our relative photometry.  They
are all about $m-M = 24.47\pm0.03$, which is greater than, but
consistent with, that of the Cepheid distance of 24.43
\cite{freedman}.  The higher distance would be consistent with some
possible blending in their sample, but not as severe as the increase
suggested by \citeN{mochejska2000}.  The measured distances are barely
in agreement with the globular cluster surface brightness fluctuation
measurement of \citeN{ajhar1996}, who found $24.56\pm0.12$.  It is
important to note that these distance results do not rule out a
greater distance because only a very limited range in the distance
parameter was allowed ($24.4\leq m-M\leq 24.5$) in order to limit the
degenerate effects of distance and reddening.  On the other hand,
these distance results have errors smaller than the range allowed, and
they are in near perfect agreement with those of \citeN{stanek1998},
who used the mean brightness of the red clump to derive a distance of
$24.471\pm0.035$
.  My distance measurements are also nearly identical to that of
\citeN{holland1998}, who measured the mean distance of the globular
cluster population to be $24.47\pm 0.07$ by fitting their red giant
branches.  Finally, my distances are consistent with those seen in
\citeN{williams2002}, whose similar analysis of HST photometry gave
distances between 24.43$\pm$0.060 and 24.505$\pm$0.042.  The mean
reddenings range from $A_V = 0.300\pm0.082$ to $A_V = 0.870\pm0.110$,
confirming the high amount of differential reddening across and
through the disk.  OB 78 has the lowest mean reddening of all the
sections, suggesting that the dust may have been moved or destroyed by
the hot stellar populations in this area.

The results of the SFH analysis for each region are shown using the
same four-paneled figure style, shown in Figures 8, 9 and 11 and
discussed in sections 4.1 - 4.3.  These figures show the observed
foreground-subtracted CMDs, the measured SFHs, the best fitting
chemical evolution histories, and the best-fitting model CMDs for each
field analyzed.

\subsection{OB 102}

One of the largest OB associations in the northern half of M31 is OB
102.  This highly elliptical association follows an outer spiral arm
that runs from OB 98 to OB 108. Because of its large size, high
H$\alpha$ luminosity and proximity to the major axis, OB 102 is an
excellent place to look for star formation propagation through the
disk.  Outside of the identification of the association and detailed
analysis of a few of its members mentioned in section 1, little is
known about this association.  In order to learn about its history, we
created three samples of stars from regions equal in area.  These
areas are outlined in Figure 1.  They are 15.8 arcmin$^2$ each.  We
selected the stars found in these regions in both B and V, and put
them through the SFH analysis described above.  There were 347, 1626
and 493 stars measured in the eastern region, OB association and
western region, respectively.  The results for each of the samples are
given in Figures 8a - 8c.

The figures show that the data are well-reproduced by this type of
analysis.  The large numbers of artificial star tests done
independently for each region provide accurate statistics for the
reproduction of completeness and photometry errors, taking into
account that many faint stars may be detected in one band and not
another.  Unfortunately, the routine is not able to accurately
reproduce the full width of the main sequence or the scattered stars
to the red of the main sequence.  The analysis technique cannot
reproduce the effects of differential reddening which broaden the main
sequence.  In addition, our limited understanding of the evolution of
high-mass stars causes problems in precisely reproducing the observed
colors and magnitudes of evolved upper main sequence stars.

The chemical evolution appears much like the results of our
experimental samples, where the input metallicity was actually
constant and solar.  This result reveals an important limitation of
the analysis technique when applied to the M31 disk: we cannot trust
the measured chemical abundance history for stars less than a few
hundred Myr in age.  

The age distribution of the eastern and western samples are identical;
however, OB 102 shows an enhanced star formation rate, not only in the
past 10 Myr, but also excess activity back to $\sim$100 Myr ago.  Here
is the first indication from these data that these large associations,
and possibly the arms of M31 themselves, have been forming stars at an
enhanced rate for the past few hundred Myr.  This result is consistent
with the results of \citeN{williams2002} that the star formation rate
in the arms has been enhanced for the past few hundred Myr; however,
an important additional conclusion can be drawn from these wider field
data: the most recent activity is indeed higher than the mean of the
past few hundred Myr in the most active regions.  This most recent
activity was difficult to assess with the limited field of view of the
HST data.  Because the most massive stars are not numerous in M31,
obtaining a significant sample requires a large field of view.  The
smaller field of HST limited the number of the most massive stars in
these samples, making it difficult to draw any conclusions about the
most recent activity and also making it difficult to compare adjacent
regions with very different populations.  Using the larger field, we
see a very recent increase in activity over a larger scale, as well as
the previously observed constant enhanced rate.

\subsection{OB 78 (NGC 206)}

OB 78 is the largest OB association in the southern half of the M31
disk.  This association has long been intriguing because of the low
nebular emission surrounding it (e.g. \citeNP{massey1986}).  One
possible explanation for the smaller amount of nebular emission in OB
78 is that the gas was blown away by massive stellar winds when the
stars were first being formed, possibly creating the H I hole observed
by \citeN{brinks1986}.  This hypothesis suggests that star formation
has recently ceased due to the exhaustion of the gas supply.  If this
explanation is correct, then there may be a noticeable difference
between the stellar populations of OB 78, where nebular emission is
low, and OB 102, where nebular emission is higher.

I performed the SFH analysis on the three regions in and around OB 78
marked in Figure 1. Each region has an area of 6.6 arcmin$^2$.  There
were 233, 887 and 193 stars measured in the northern region, OB
association and southern region, respectively.  The results of the SFH
analysis are shown in Figures 9a - 9c.  The stellar populations of
these three regions bare a striking similarity to those of their
counterparts surrounding OB 102.  A nearly constant low star formation
rate is measured for the regions next to the OB association.  One data
point, at 200 Myr for the region north of OB 78, deviates from this
description; however this point is more likely an outlier due to
slight errors in the completeness measurements at the faint-end limit
of the data.  In the case of OB 78, the association itself shows a
sharp increase in star formation rate back to 30 Myr ago.  Before 30
Myr ago, the rate was comparable to that now observed for the adjacent
regions.  According to this analysis, OB 102 and OB 78 contain stellar
populations with one significant difference: the significant increase
in star formation formation in OB 78 began $\sim$30 Myr ago while such
an increase did not occur in OB 102 until $\sim$10 Myr ago.  Perhaps
this earlier increase in activity in OB 78 cleared a significant
amount of the diffuse gas from the region so that less emission is
seen from the most recently formed stars.  I roughly measured the
H$\alpha$ luminosities of OB 102 and OB 78 from the data of
\citeN{winkler1995} using circular apertures of radius 50 arcsec for
each association.  These rough luminosities are $\sim1.0 \times
10^{38} erg/s$ and $\sim3.8 \times 10^{37} erg/s$ respectively.  I
show below that OB 78 appears to have more of the youngest stars per
unit area than OB 102; however, its H$\alpha$ emission is smaller than
OB 102 by more than a factor of 2. Such a result is important to
understand when applying conversions of H$\alpha$ luminosity to star
formation rate for distant galaxies; regions with H$\alpha$
luminosities varying by more than a factor of 2 may have nearly
identical current star formation rates.

I tested the validity of the difference between the populations of OB
102 and OB 78 seen by the SFH analysis by looking for a difference
between the luminosity functions of the two regions.  The luminosity
functions of the two regions were measured by taking a sample of the
stars from each region between -0.4 $\leq$ B-V $\leq$ 0.2 and 16
$\leq$ V $\leq$ 23.  These samples were binned by 0.2 in B-V color and
0.3 in V magnitude, and each bin was corrected for completeness
determined from the artificial star tests and for extinction
determined from the MATCH analysis.  Finally, the total
completeness-corrected number of stars at each V magnitude bin was
computed.  After adopting a distance modulus to M31 of 24.47 and
normalizing each bin to an area of 1 arcmin$^2$, the luminosity
functions of the two regions are shown in Figure 10.

The faint ends of the two luminosity functions are different, showing
why the analysis routine measured a high star formation rate 200 Myr
ago for OB 78.  The validity of this difference is difficult to assess
because it is strongly dependent upon precise knowledge of the
completeness at the detection limit of the data.  At the same time,
There is a significant difference in the shape of the luminosity
functions brightward of $M_V \sim -6.5$.  This difference supports the
result from the SFH analysis, showing that the age distributions of
these populations differ.  OB 78 contains more stars between 10 and 30
Myr old, whose progenitors likely blew out much of the diffuse gas in
the region.  This scenario also explains the low extinction of OB 78,
which is consistent with being entirely due to dust in our Galaxy.
Apparently, the dust was blown out along with the gas.  It appears as
though in another 20 Myr OB 102 will have an H$\alpha$ luminosity more
like that of OB 78.

The results do not show any obvious evidence for the propagation of
star formation due to spiral density wave passage suggested by
\citeN{magnier1997a} and
\citeN{williams2001a}.  
One possible signature for a density wave passage would be a significantly
larger population of stars with ages of $\sim$10-100 Myr on one side
of the current density wave location than on the other.  Such a
signature was seen in the Cepheid age distribution of the southern
half of the disk, on a very large size scale
\cite{magnier1997a}.  The areas near OB 78 and OB 102 show nearly
identical stellar populations on either side of the associations.  We
do not see this signature for spiral wave passage on the scale of our
chosen regions.  This result does not rule out the conclusions of
previous work on the southern half of the disk, where a signature for
spiral wave propagation was seen on a much larger size scale.  Our
study does not rule out the possibility that there is a difference
between the populations on either side of the association on larger
scales, better probed by the ages of the star clusters and Cepheids in
the region.  The analysis is hampered on such large size scales by the
varying completeness and increased range of extinction in the data.
Indeed, the fact that OB 78 shows no signs of enhanced star formation
before 30 Myr ago is consistent with the idea that the density wave
responsible for this large OB association was previously creating
stars in a different region of the disk.

\subsection{The Northeast Spiral Arm}

The northeast spiral arm of M31 (also known as N4 and N5;
\citeNP{baade1963}) is the most well-studied region of the M31 disk.
The high interest is likely due to the star formation activity in this
region which produces strong H$\alpha$ emission.  The region contains
hundreds of discrete nebulae \cite{walterbos1992}, including hundreds
of supernova remnants (e.g. \citeNP{braun1993}, \citeNP{magnier1995},
\citeNP{williams1995}) and H II regions (e.g. \citeNP{walterbos1999}).
It contains dozens of open clusters (\citeNP{hodge1979},
\citeNP{williams2001b}) and OB associations \cite{efremov1987}, and it
shows evidence for having been active for the past several hundred Myr
\cite{williams2002}.  Clearly there is star formation occurring is
this well-studied spiral arm.  Because it is so active and has such
sharp boundaries in the H$\alpha$ images, it is a good region to look
for patterns of star formation around spiral arms using the SFH
analysis technique.  I performed SFH measurements on three regions
associated with this arm.  These regions are shown in Figure 1: one to
the east of the arm, one on the arm, and one to the west of the arm.
The regions each cover an area of 27.8 arcmin$^2$.  There were 345,
1350 and 304 stars measured in the eastern region, arm and western
region, respectively.  The results from each of these regions are
shown in Figures 11a - 11c.

Again, the results show that the arm has been more active than the
surroundings for at least 100 Myr, and until very recently, the star
formation rate was constant.  This constant rate over the past
$\sim$100 Myr is a nice confirmation of the conclusion of
\citeN{williams2002} from HST data in this region of the galaxy.
Those deeper, but smaller, fields revealed that the star formation
rate has been fairly constant and high in the spiral arms over the
past few hundred Myr.  These larger fields now reveal a factor of 3
rise in the star formation rate in the past several Myr.  This jump
was not possible to measure conclusively with the smaller fields
because the sampling of the youngest stars was inadequate over such
small areas.  Once again, there is no evidence that the star formation
has propagated from either side of the current position of the arm.
Neither side of the arm shows a significantly higher number of young
stars than the other, which would be an indication of the recent
passage of a density wave.  The surrounding regions have had low,
relatively constant star formation for the past 100 Myr.

\section{Conclusions and Future Work}


I have measured preliminary B and V photometry for resolved stars in
the LGS data set for the M31 disk.  From this photometry I have
determined SFHs for three of the largest active regions of the disk
and the areas surrounding those regions in order to understand the
difference in nebular emission from the northern and southern halves
of the disk, to look for consistency with previous analyses of HST
data in overlapping regions and to look for patterns of star formation
in these regions of the disk.  In particular, I looked for evidence of
star formation propagation through the disk near the spiral arms.

None of the three regions selected showed unambiguous evidence for
star formation propagation, such as an over-abundance of stars with
ages 10-100 Myr on one side of the currently active regions.  On the
other hand, the regions all showed common characteristics.  OB 78, OB
102, and the northeast spiral arm all show a recent increase in
activity from the past average.  This increase began earlier in OB 78
than in the other two regions, which may explain the weaker nebular
emission measured in this region.  The areas adjacent to each region
show no effects of being so close to star formation activity.  Their
evolution appears to have been unaffected over the past 100 Myr that
these areas have had higher star formation rates.  Similar results
were seen in \citeN{williams2002}.  This evidence leads to the
conclusion that active regions have fairly long lifetimes: perhaps up
to several hundred Myr.  During this lifetime, they appear to have
factor of $\sim$2 enhancements in activity on shorter timescales.
These enhancements will likely increase the nebular emission if the
diffuse gas has not been blown away by previous star formation;
however, if the high star formation rate continues for more than
$\sim$20 Myr, much of the diffuse gas may be blown away from the
regions and the nebular emission can decrease even though the star
formation rate is the same.


The LGS data set \cite{massey2001} can be applied to many more
scientific questions, particularly once the data are fully calibrated.
When SFHs of each region in the M31 disk are measured individually, it
is possible to overcome the complications imposed by differential
reddening: an effect intensified in M31 by its high inclination angle.
Current plans include systematically determining the recent SFH as a
function of position over the entire disk to about 4 arcminute
resolution, as well as learning more about the older stellar
populations through an analysis of the I band data, which contains at
least five times the number of stars recovered in the B band.  If the
crowding issues prove to be manageable, SFHs for the older populations
of the disk as a function of position will be measured from the large
sample.

\subsection{Acknowledgments}

I thank the LGS team for making this project possible by supplying the
data.  I especially thank Phil Massey for his work in observing,
preparing and providing the images for the photometric measurements.
Finally, I thank Paul Hodge for mentoring the project.



\clearpage

\begin{deluxetable}{ccccc}
\tablenum{1}
\small
\tablecaption{Zero points determined for each M31 field from previous catalogs}
\tablehead{
\colhead{\bf{Field}} &
\colhead{\bf{B Zero}} &
\colhead{\bf{B Stars}} &
\colhead{\bf{V Zero}}  &
\colhead{\bf{V Stars}} 
}
\startdata
F1 & 4.08$\pm$0.18 & 1773 & 4.36$\pm$0.18 & 1651\nl
F2 & 3.95$\pm$0.16 & 488 &  3.80$\pm$0.18 & 488\nl
F3 & 4.12$\pm$0.14 & 1294 & 4.39$\pm$0.12 & 3004\nl
F8 & 4.06$\pm$0.22 & 366 & 4.37$\pm$0.20 & 379\nl
F9 & 3.94$\pm$0.26  & 155 & 4.32$\pm$0.18 & 148\nl
F10 & 3.91$\pm$0.26 & 119 & 4.34$\pm$0.18 & 68\nl
\enddata
\end{deluxetable}

\begin{deluxetable}{ccc}
\tablenum{2}
\small
\tablecaption{Distances and Reddening Values Determined for Each Field by MATCH}
\tablehead{
\colhead{\bf{Region}} &
\colhead{\bf{m-M}} &
\colhead{\bf{$A_V$}}
}
\startdata
East of OB 102 & 24.450$\pm$0.041 & 0.611$\pm$0.099\nl
OB 102 & 24.475$\pm$0.025 & 0.500$\pm$0.082\nl
West of OB 102 & 24.475$\pm$0.025 & 0.500$\pm$0.082\nl
East of NE Arm & 24.460$\pm$0.037 & 0.870$\pm$0.110\nl
NE Arm & 24.467$\pm$0.033 & 0.567$\pm$0.115\nl
West of NE Arm & 24.475$\pm$0.025 & 0.450$\pm$0.112\nl
North of OB 78 & 24.475$\pm$0.025 & 0.450$\pm$0.171\nl
OB 78 & 24.475$\pm$0.025 & 0.300$\pm$0.082\nl
South of OB 78 & 24.478$\pm$0.025 & 0.378$\pm$0.131\nl
\enddata
\end{deluxetable}
\clearpage

\begin{figure}
\figurenum{1} 
\centerline{\psfig{file=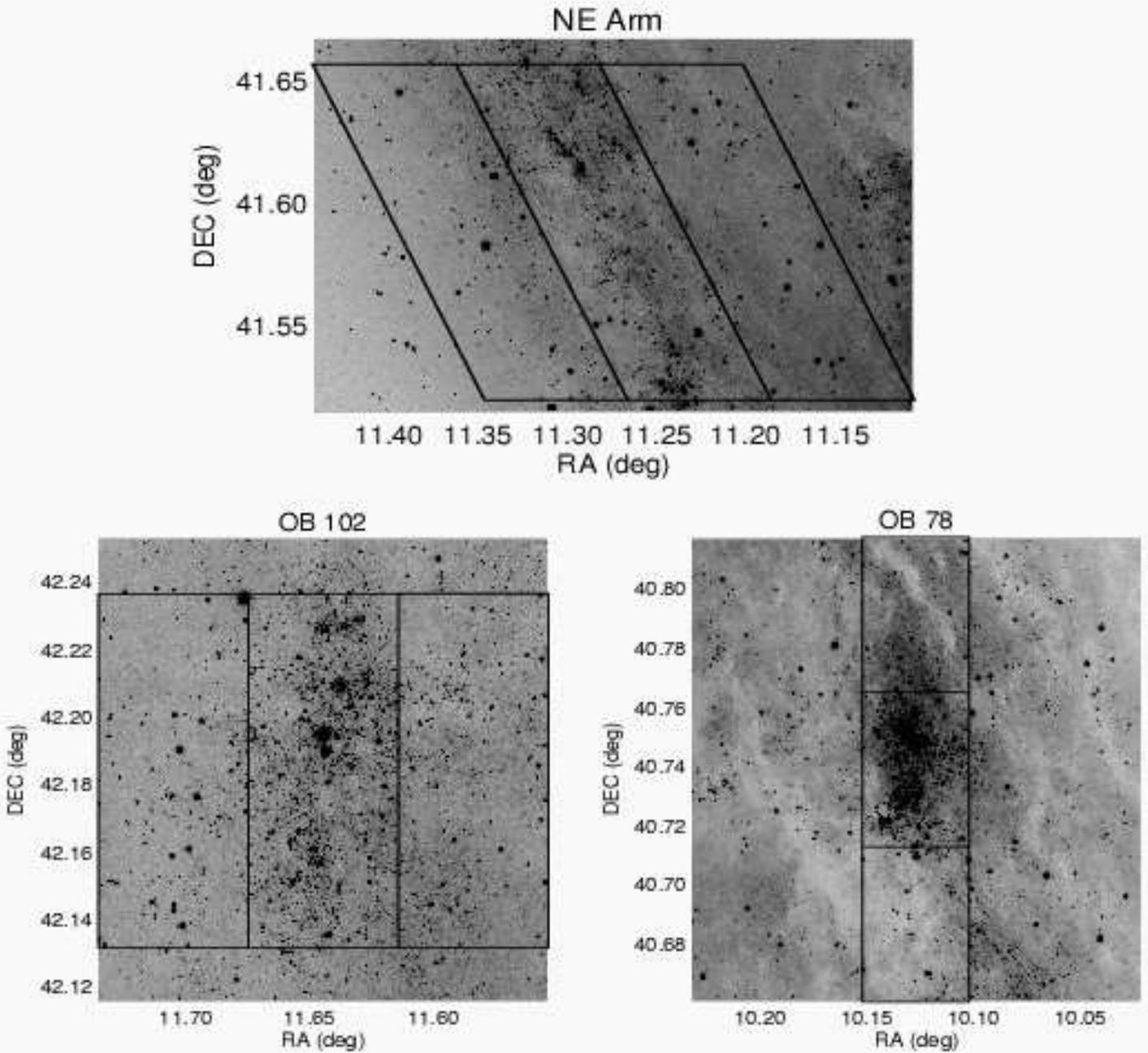,height=8.0in,angle=0}}
\caption[Ground-based fields taken for the LGS survey]{(a)Stacked B
band images of the three areas of M31 under study.  Each panel shows
the three regions of the area whose SFHs were measured.  The top panel
shows the NE spiral arm, taken from field F3.  The lower left panel
shows OB 102, taken from field F2. The lower right whows OB 78, taken
from field F8.}
\end{figure}

\begin{figure}
\figurenum{2} 
\centerline{\psfig{file=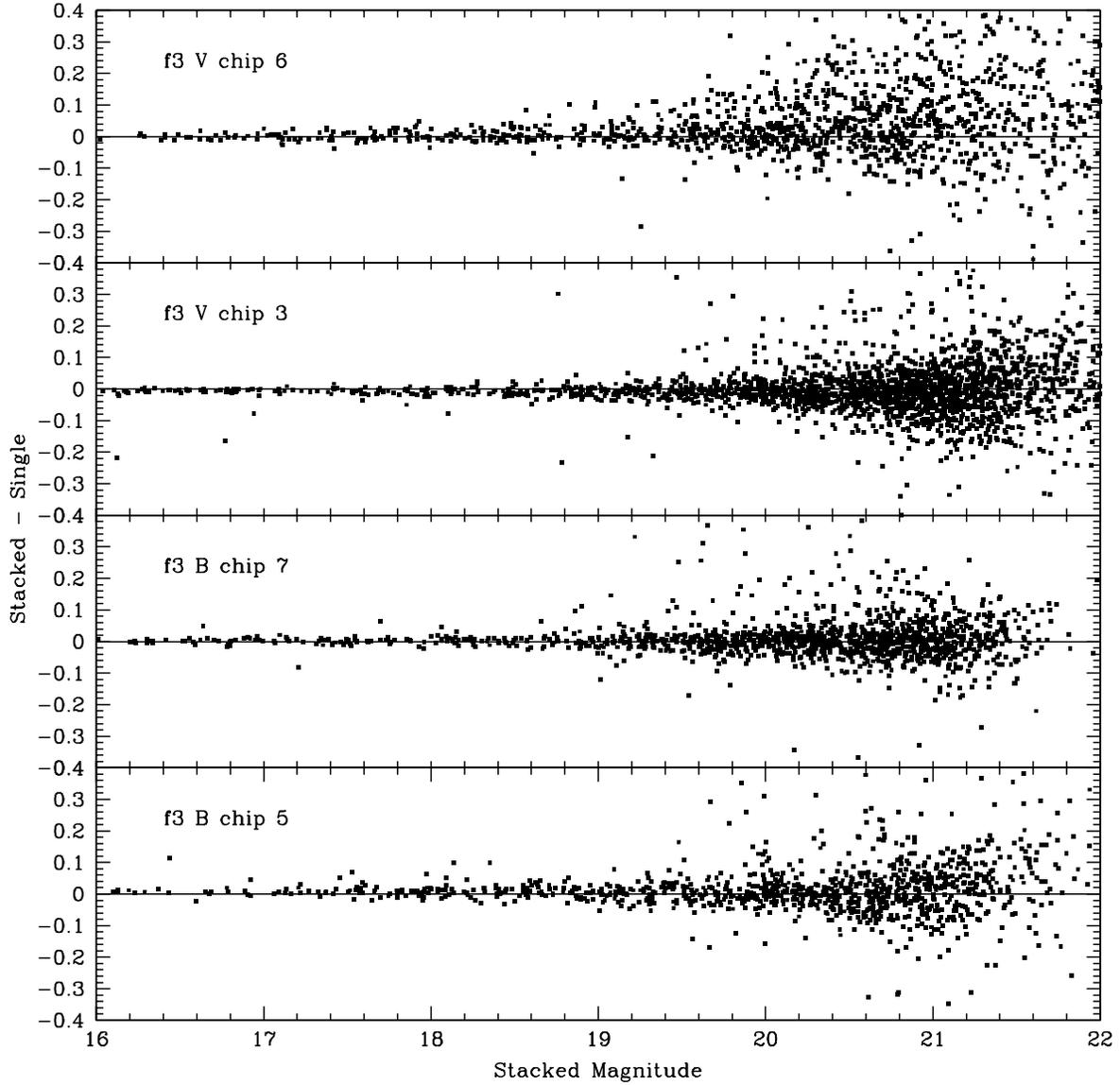,height=6.0in,angle=0}}
\caption[Comparing stacked and individual exposure PSF U band photometry]{Shown are the differences of PSF magnitudes obtained from an individual exposure and those obtained from a stacked image in the B and V bands.  The top two panels show the residuals vs. the final stacked magnitude for two randomly chosen chips in the B band.  The bottom two panels show the same quantities for the V band.} 
\end{figure}

\begin{figure}
\figurenum{3} 
\centerline{\psfig{file=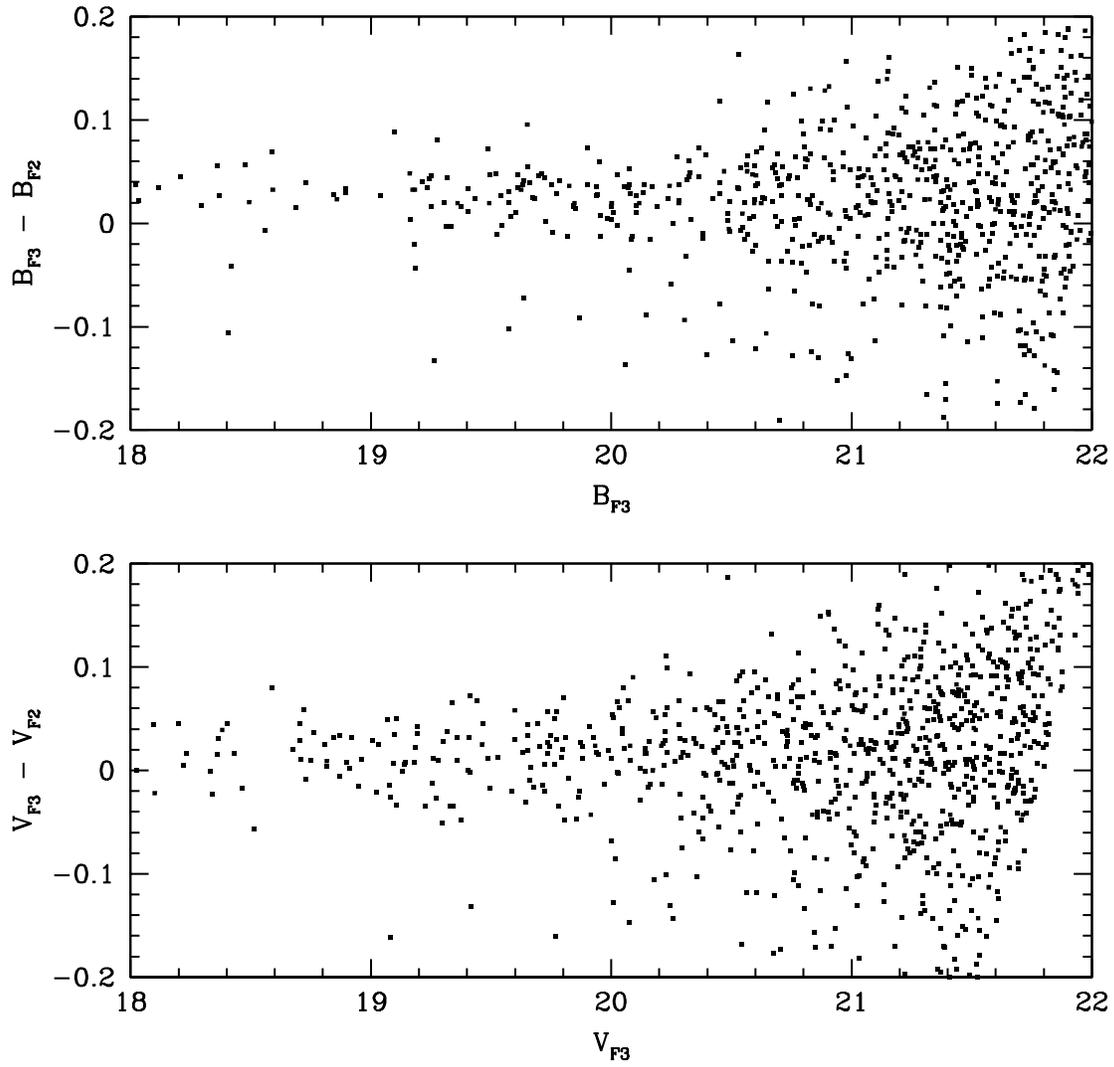,height=6.0in,angle=0}}
\caption[Comparing multiple measurements of stars]{The residuals of the photometry of the same stars measured as measured in F2 and F3 are plotted vs. their F3 magnitude.  The independent B and V band measurements show good agreement.} 
\end{figure}

\begin{figure}
\figurenum{4} 
\centerline{\psfig{file=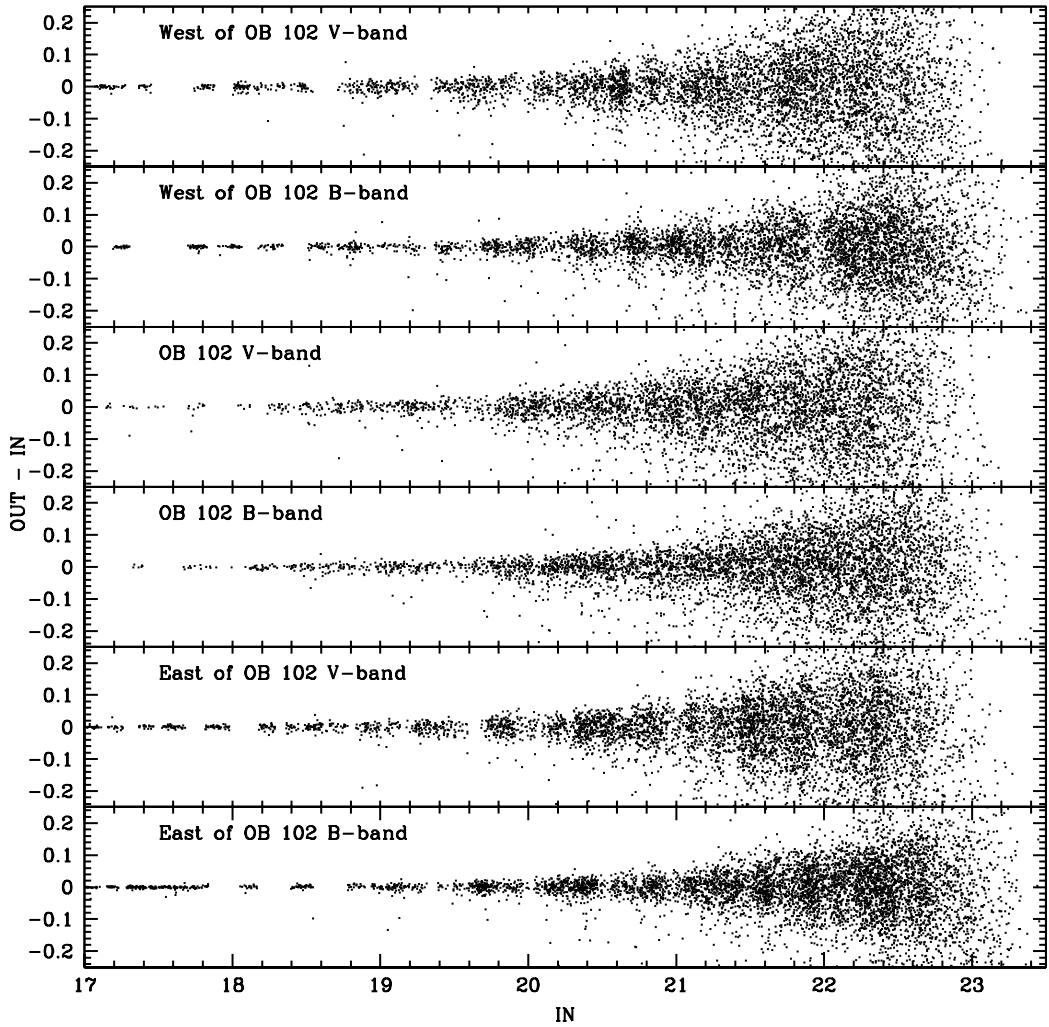,height=6.0in,angle=0}}
\caption[Artificial star tests around OB 102]{The residuals of the measured photometry of the stars I added to the image sections are plotted against the input magnitudes.  Notice the excellent relative photometry as well as the differences in completeness at the faint end in the different regions.} 
\end{figure}

\begin{figure}
\figurenum{5} 
\centerline{\psfig{file=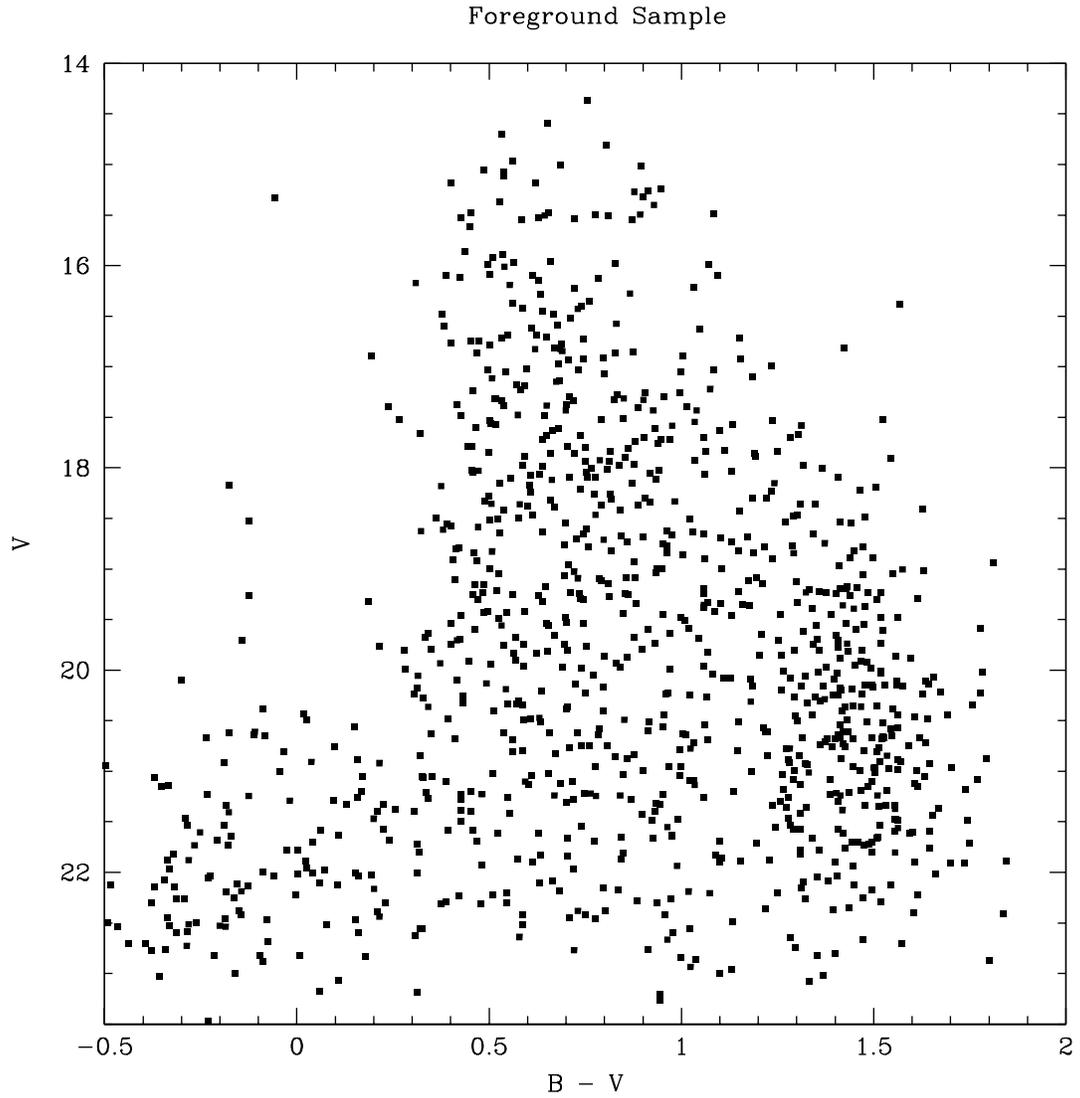,height=6.0in,angle=0}}
\caption[Foreground star sample]{Sample of stars taken from the portions of the data most distant from the inner disk of M31.  These areas are marked in Figures 1a and 1f.  They were used to estimate the foreground contamination in the areas of interest in the disk.} 
\end{figure}

\clearpage

\begin{figure}
\figurenum{6} 
\centerline{\psfig{file=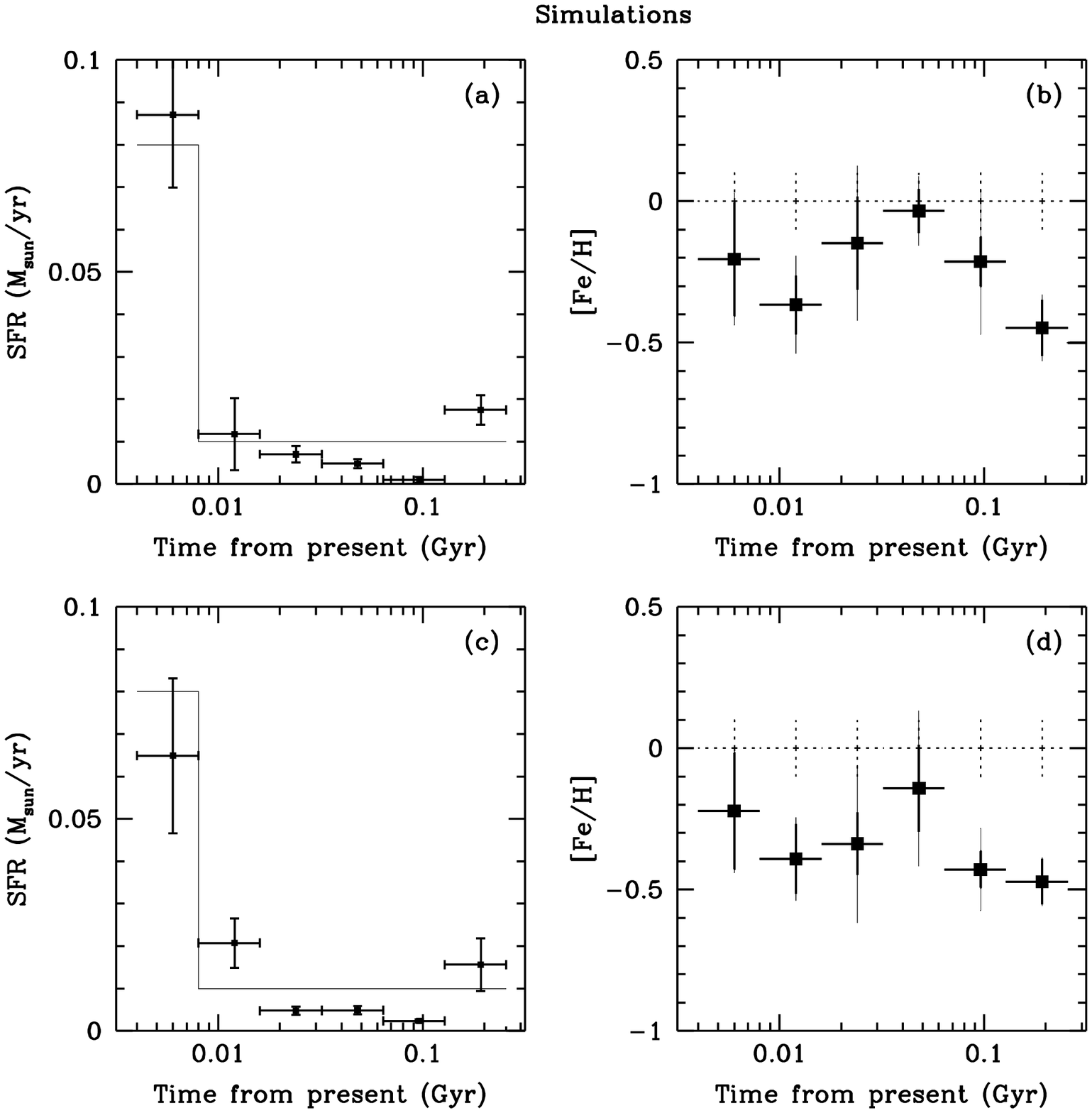,height=6.0in,angle=0}}
\caption[Synthetic sample tests of SFH analysis routine]{The results
from the synthetic star sample experiment are shown. (a) Points with
error bars show the SFH determined for the constant star formation
rate (shown with the solid line) put into the analysis routine. (b)
Solid square points with error bars show the abundance history
determined by the analysis routine. Heavy error bars mark the
metallicity range for the time period, and the light error bars show
how the measured errors of the mean metal abundance for the time
period could shift the metallicity range.  The input abundance history
is shown by the dashed line.  Dashed errors mark the input metallicity
range.  (c and d) Same as a and b, but the stars in the field were given unique reddening values, $A_V$ ranging from +0.4 to -0.4 from the mean value.}
\end{figure}

\begin{figure}
\figurenum{7} 
\centerline{\psfig{file=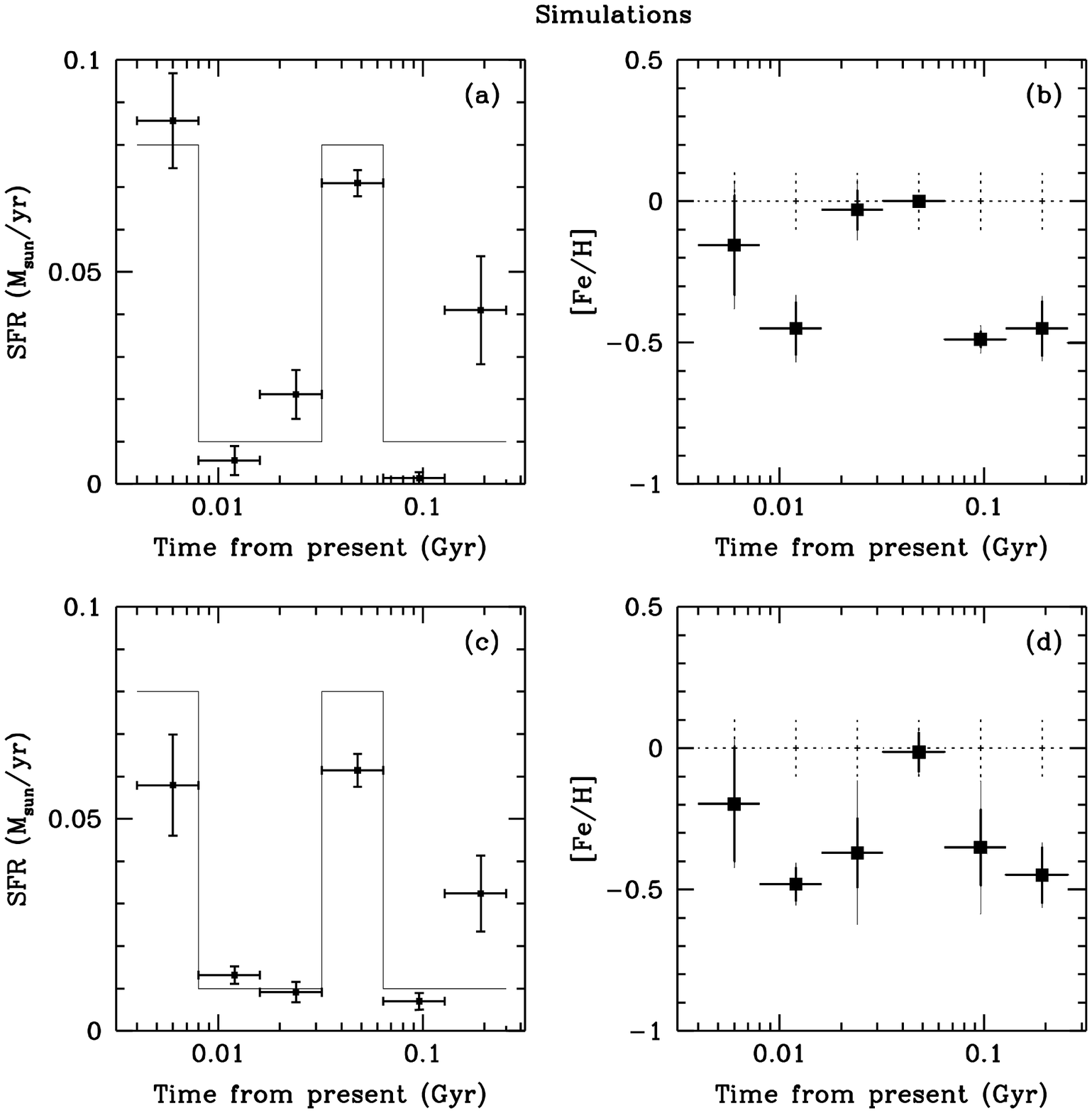,height=6.0in,angle=0}}
\caption[synthetic sample tests of SFH analysis routine]{The results
from a second synthetic star sample experiment are shown.  (a) Points with
error bars show the SFH determined for the constant star formation
rate (shown with the solid line) put into the analysis routine. (b)
Solid square points with error bars show the abundance history
determined by the analysis routine. Heavy error bars mark the
metallicity range for the time period, and the light error bars show
how the measured errors of the mean metal abundance for the time
period could shift the metallicity range.  The input abundance history
is shown by the dashed line.  Dashed errors mark the input metallicity
range.  (c and d) Same as a and b, but the stars in the field were given unique reddening values, $A_V$ ranging from +0.4 to -0.4 from the mean value.} 
\end{figure}

\clearpage

\begin{figure}
\figurenum{8} 
\centerline{\psfig{file=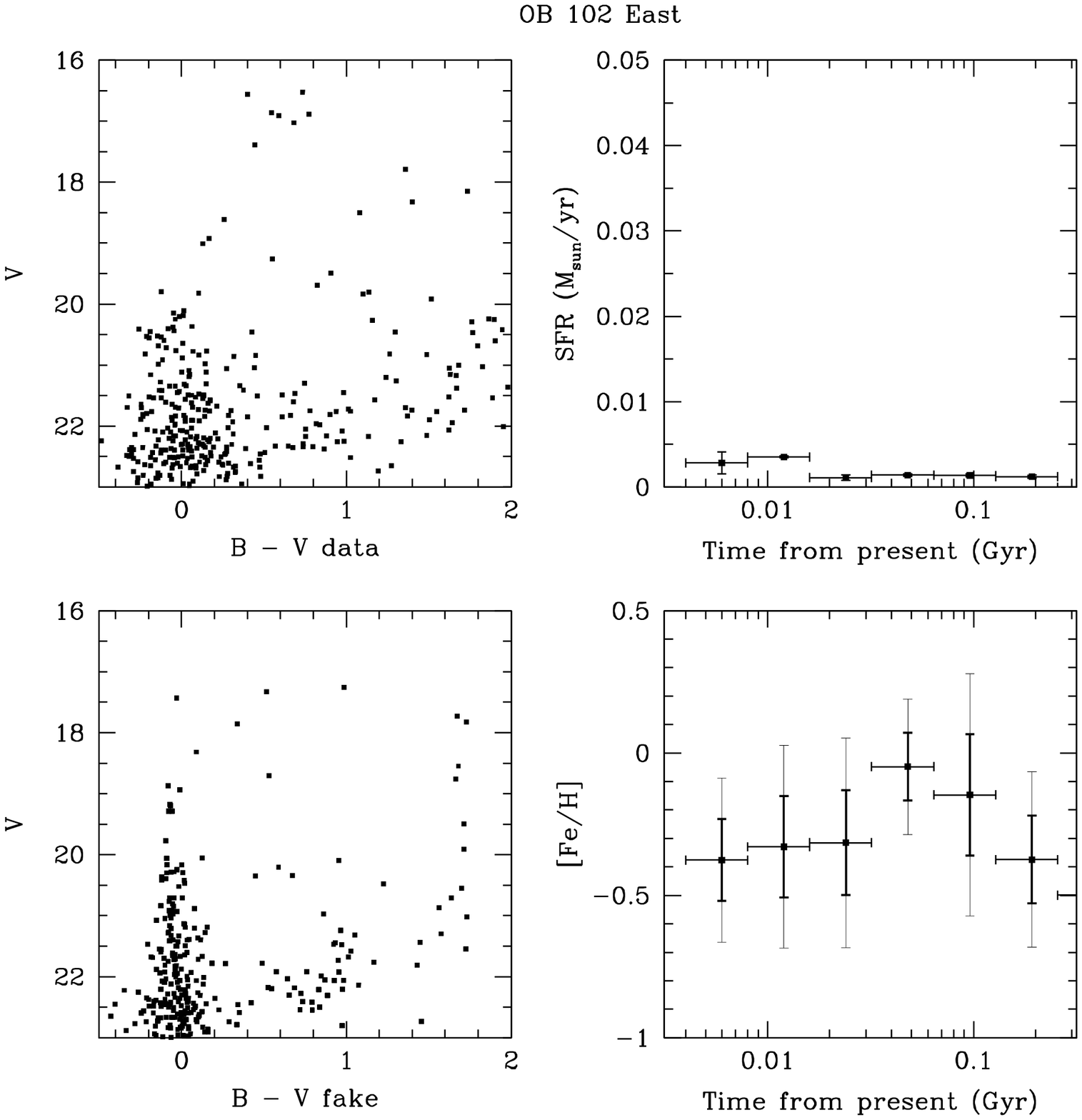,height=6.0in,angle=0}}
\caption[Star formation history of the region around OB 102]{(a) The
SFH of the region east of OB 102 is shown.  The upper-left panel shows
the observed CMD with the foreground removed.  The upper-right panel
gives the best-fitting SFH for the field.  The lower-right panel shows
the best fitting chemical evolution for the field, but experiments
show that this is not reliable. (Heavy error bars show the measured
metallicity range; lighter error bars show how the measured errors
could shift that range.)  Finally, the lower-left panel provides the
synthetic CMD created by applying the SFH in the upper-right to the
stellar evolution models of \citeN{girardi2000} and simulating
photometry and completeness errors measured with artificial star
tests.}
\end{figure}

\begin{figure}
\figurenum{8} 
\centerline{\psfig{file=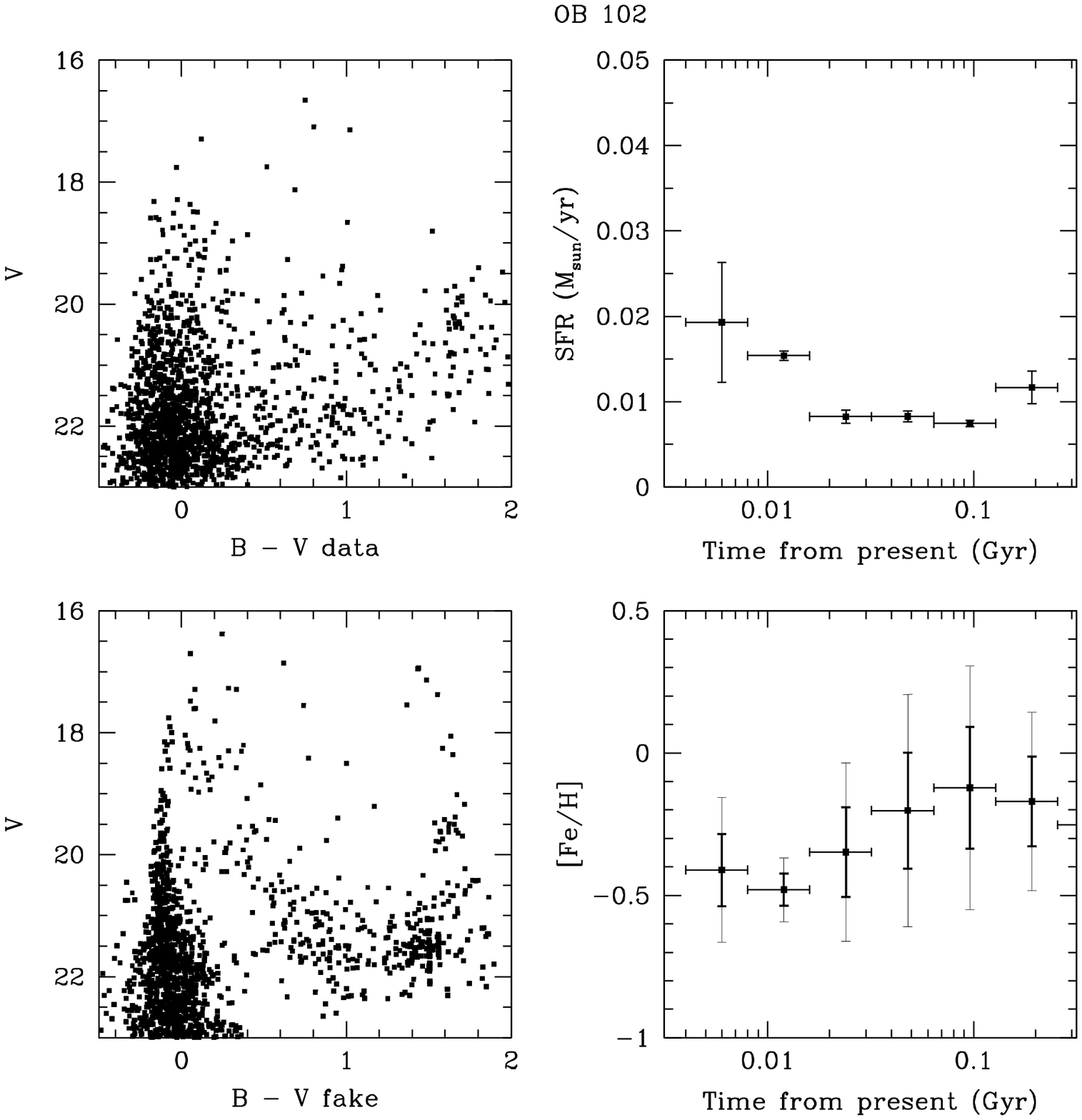,height=6.0in,angle=0}}
\caption{(b) The SFH of OB 102 is shown.  The history reveals that the associations has been forming more stars than the surroundings for the past $\sim$100 Myr, as well as a very recent increase in the star formation rate.}
\end{figure}

\begin{figure}
\figurenum{8} 
\centerline{\psfig{file=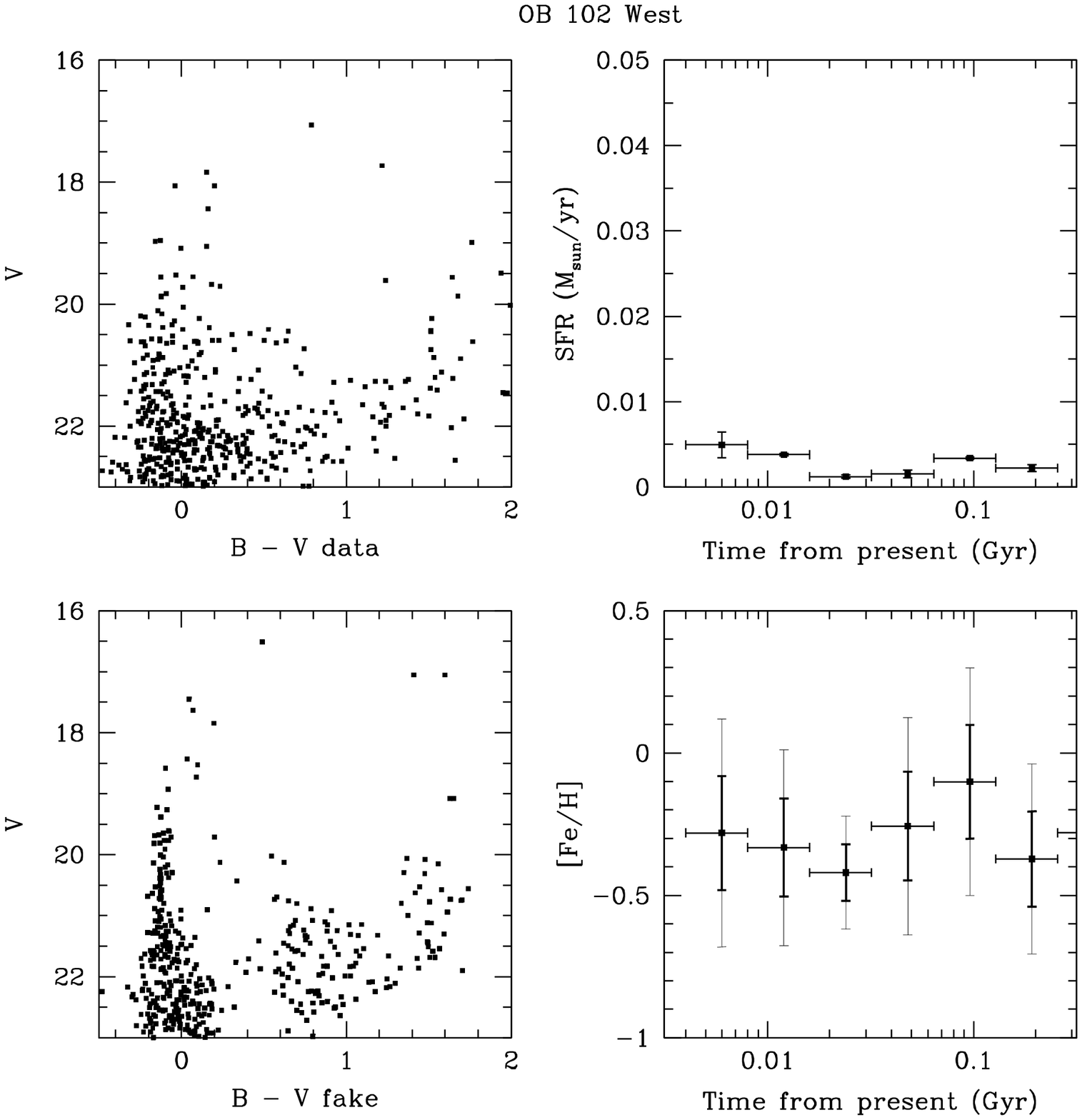,height=6.0in,angle=0}}
\caption{(c) The SFH of the region west of OB 102 is shown, and it is statistically equivalent to the region east of the association.}
\end{figure}

\begin{figure}
\figurenum{9} 
\centerline{\psfig{file=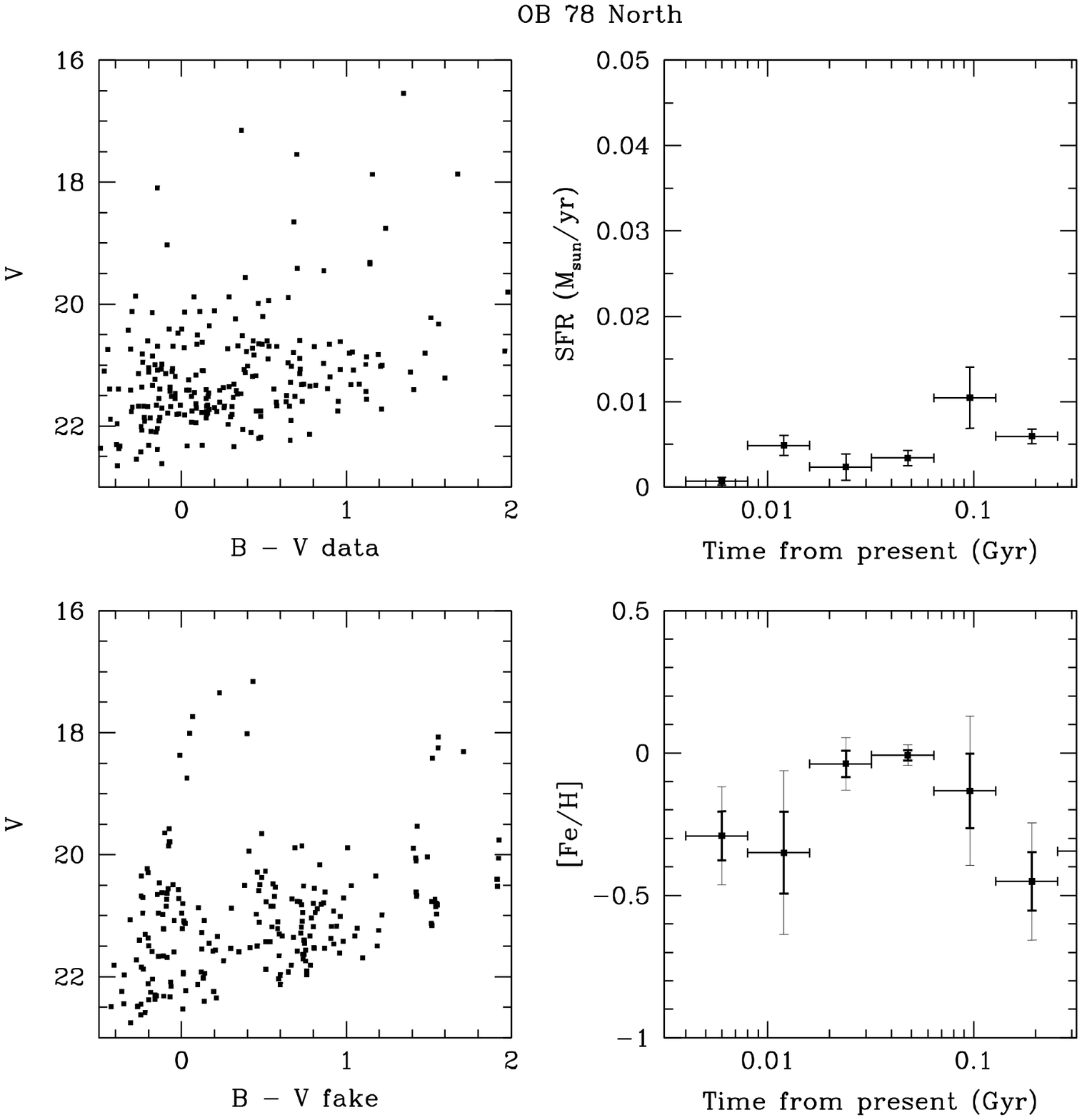,height=6.0in,angle=0}}
\caption[Star formation history of the region around OB 78]{(a) The
SFH of the region north of OB 78 is shown.  The upper-left panel shows
the observed CMD with the foreground removed.  The upper-right panel
gives the best-fitting SFH for the field.  It shows a small burst of
activity at 100 Myr, the reality of which is questionable.  The
lower-right panel shows the best fitting chemical evolution for the
field, but experiments show that this is not reliable. (Heavy error
bars show the measured metallicity range; lighter error bars show how
the measured errors could shift that range.)  Finally, the lower-left
panel provides the synthetic CMD created by applying the SFH in the
upper-right to the stellar evolution models of \citeN{girardi2000} and
simulating photometry and completeness errors measured with artificial
star tests..}
\end{figure}

\begin{figure}
\figurenum{9} 
\centerline{\psfig{file=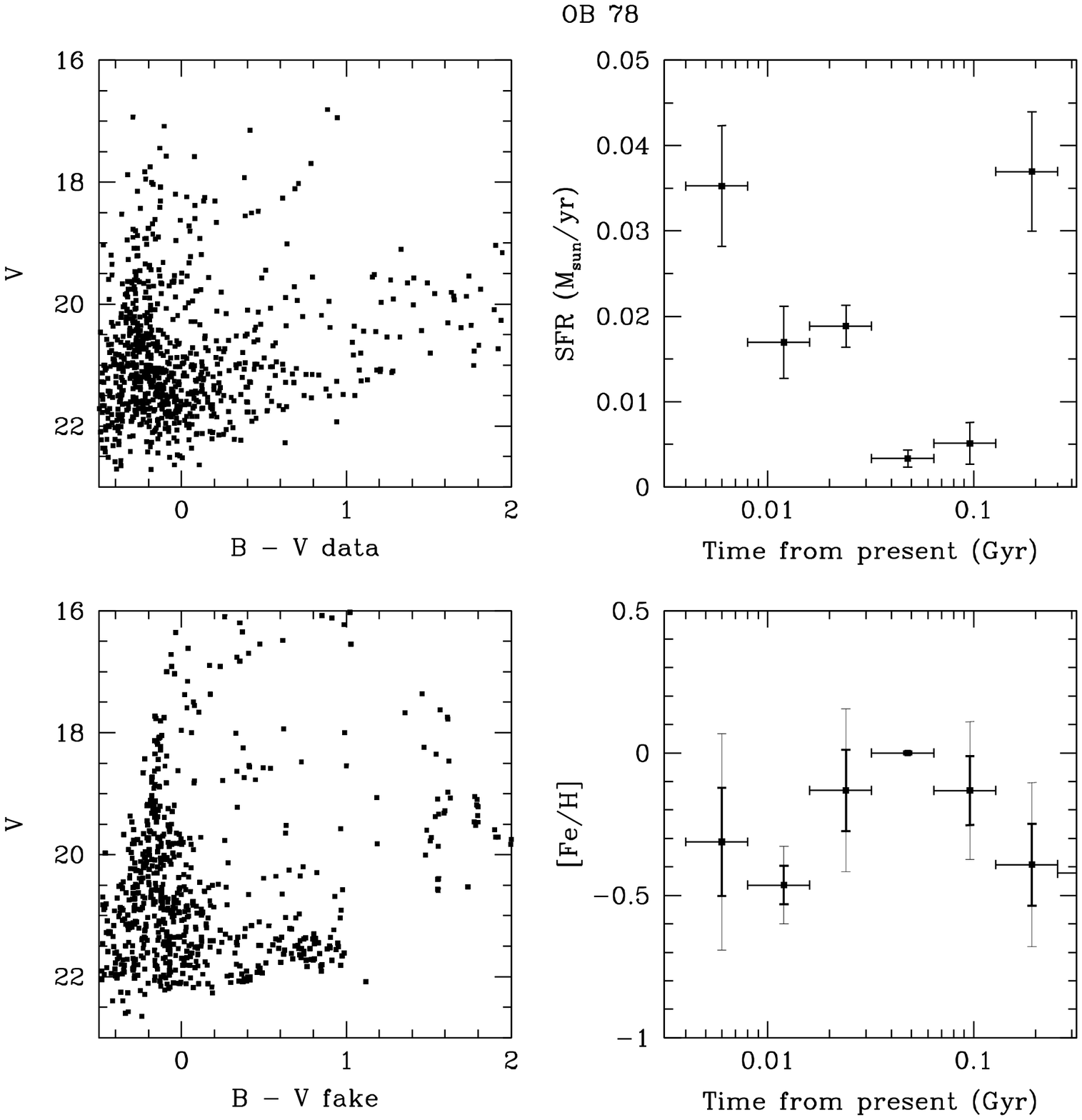,height=6.0in,angle=0}}
\caption{(b) The SFH of OB 102 is shown.  The history reveals that the associations has been forming more stars than the surroundings for the past $\sim$100 Myr, as well as a very recent increase in the star formation rate.}
\end{figure}

\begin{figure}
\figurenum{9} 
\centerline{\psfig{file=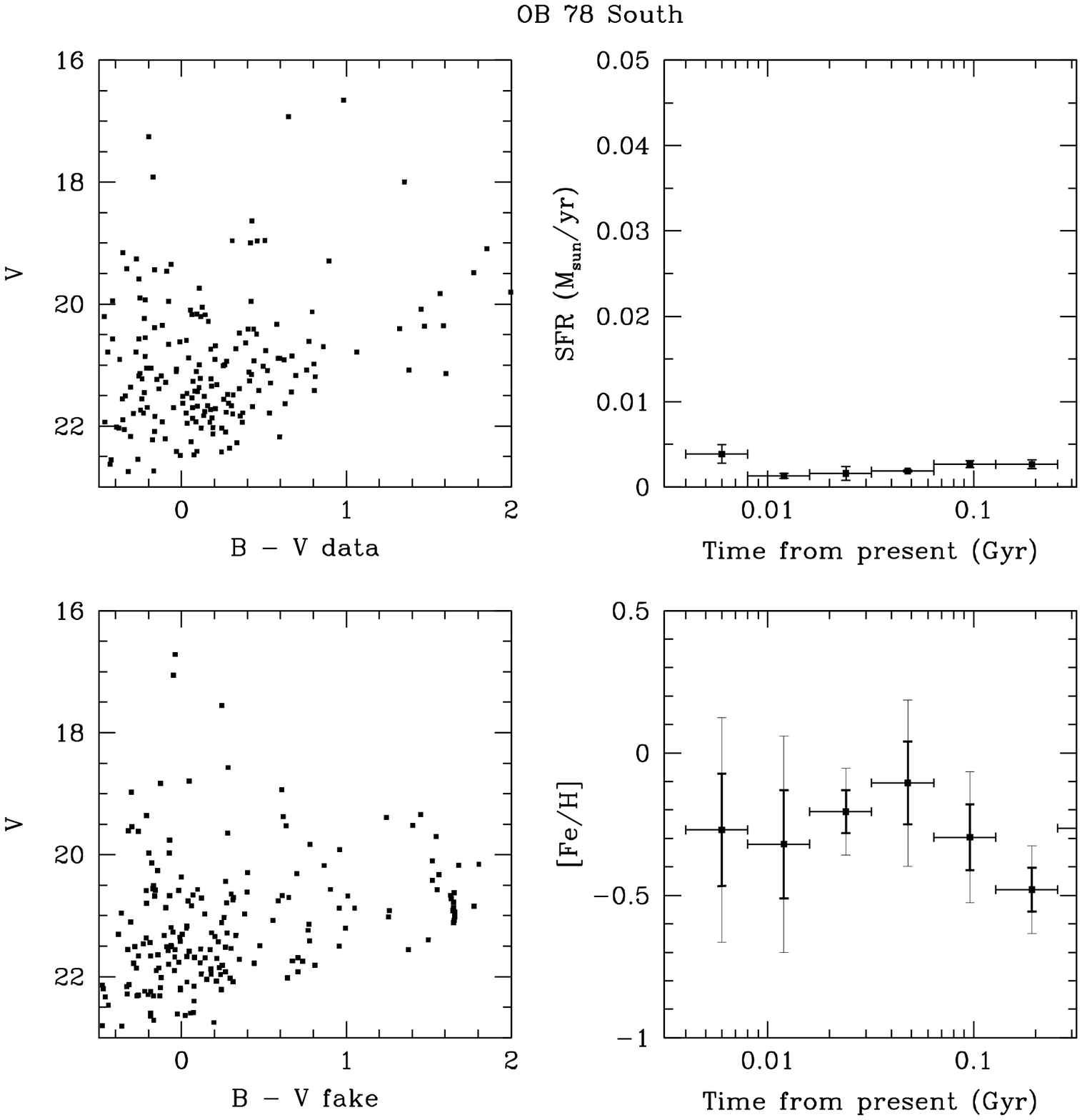,height=6.0in,angle=0}}
\caption{(c) The SFH of the region south of OB 78 is shown, and it similar to the region east of the association.}
\end{figure}

\begin{figure}
\figurenum{10} 
\centerline{\psfig{file=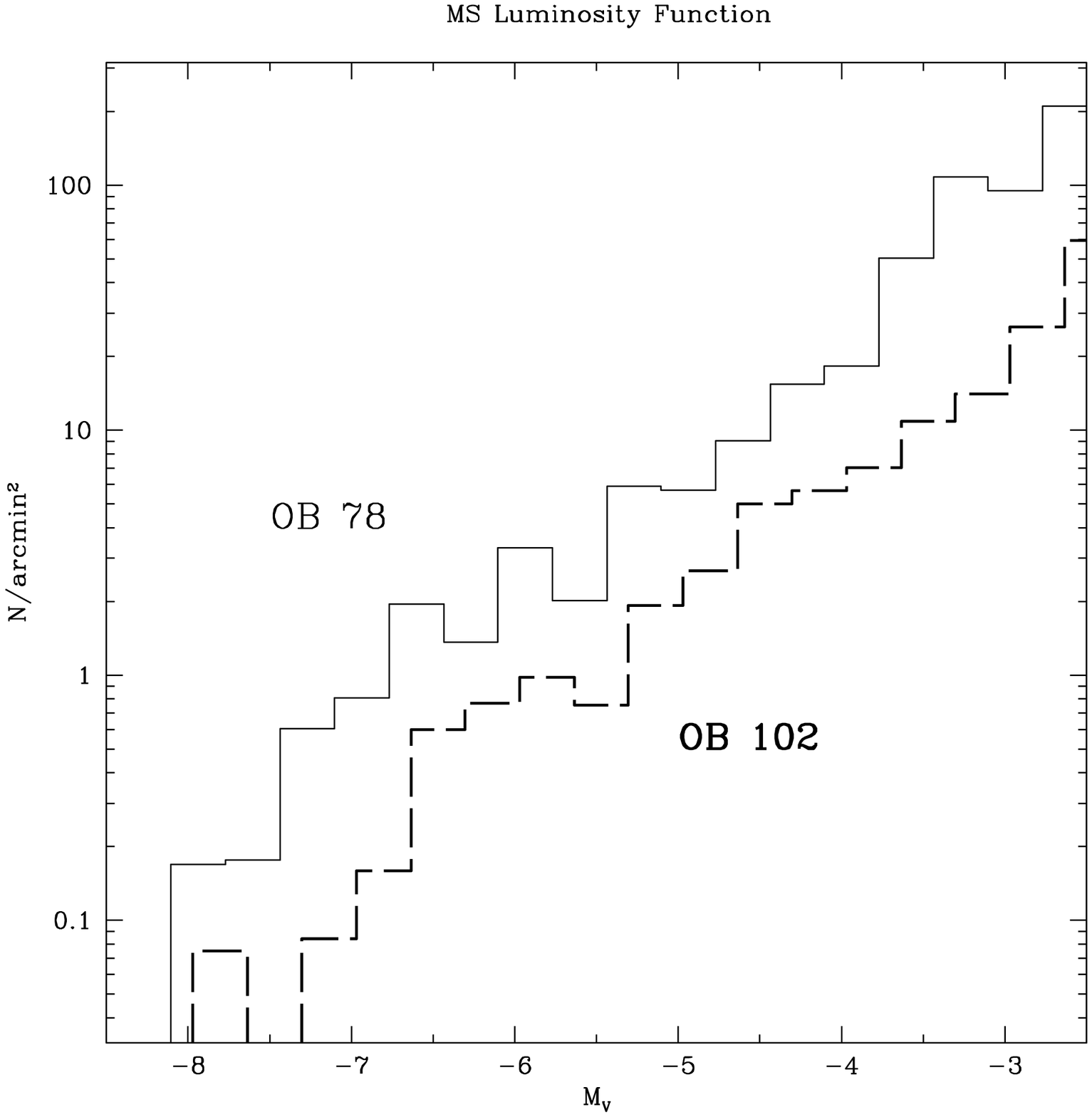,height=6.0in,angle=0}}
\caption[Main sequence luminosity functions of OB 102 and OB 78]{A comparison of the LFs of OB 102 and OB 78.  Fractional star numbers are the result of the normalization to 1 arcmin$^2$.  A difference in the bright end is clearly seen, revealing the accuracy of the SFH determined by the analysis.  This difference shows that OB 78 has had a high star formation rate for about 20 Myr longer than OB 102.}
\end{figure}

\begin{figure}
\figurenum{11} 
\centerline{\psfig{file=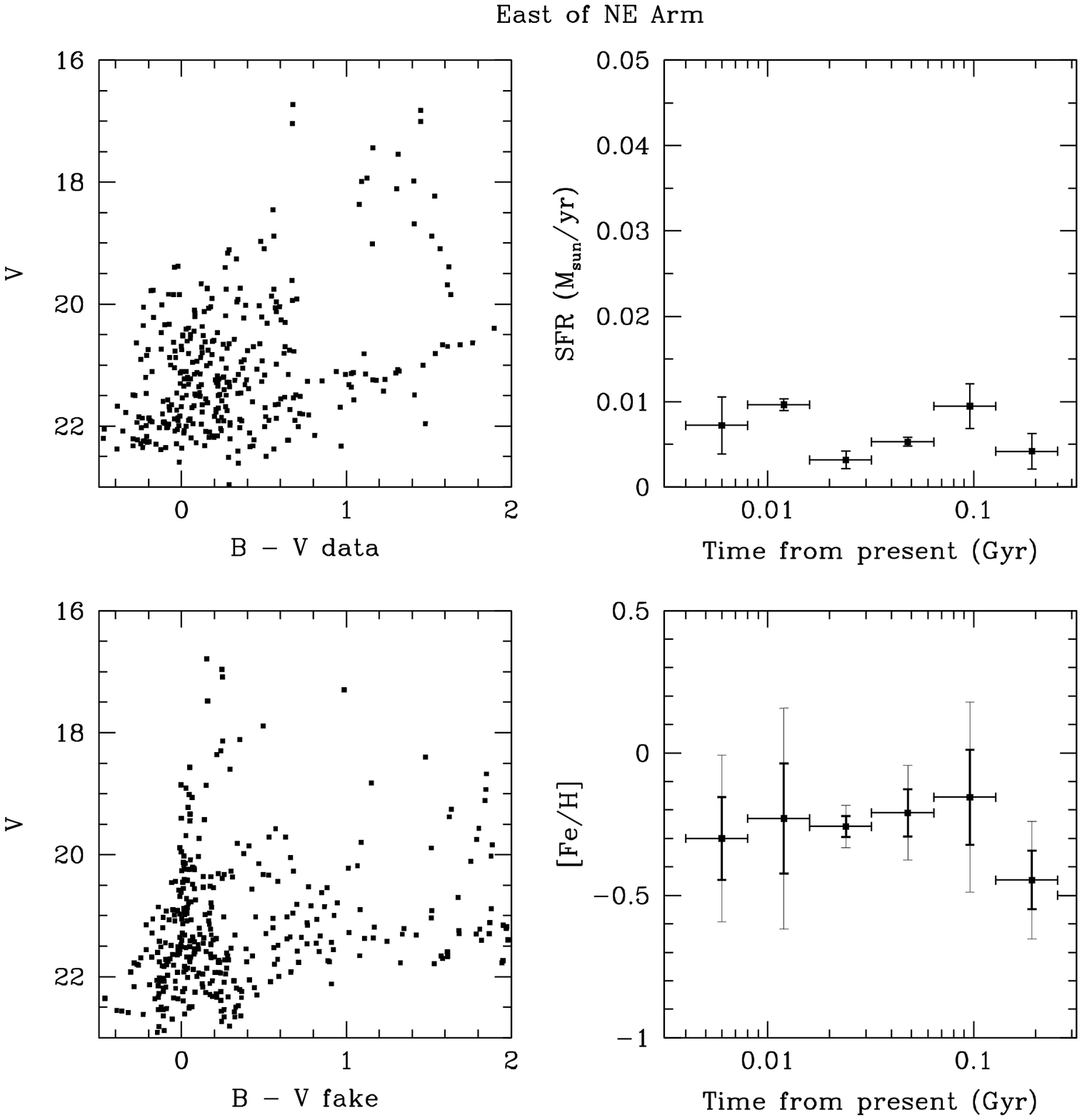,height=6.0in,angle=0}}
\caption[Star formation history of the region around the northeast
spiral arm]{(a) The SFH of the region east of the arm is shown.  The
upper-left panel shows the observed CMD with the foreground removed.
The upper-right panel gives the best-fitting SFH for the field.  The
lower-right panel shows the best fitting chemical evolution for the
field, but experiments show that this is not reliable. (Heavy error
bars show the measured metallicity range; lighter error bars show how
the measured errors could shift that range.)  Finally, the lower-left
panel provides the synthetic CMD created by applying the SFH in the
upper-right to the stellar evolution models of \citeN{girardi2000} and
simulating photometry and completeness errors measured with artificial
star tests.}
\end{figure}

\begin{figure}
\figurenum{11} 
\centerline{\psfig{file=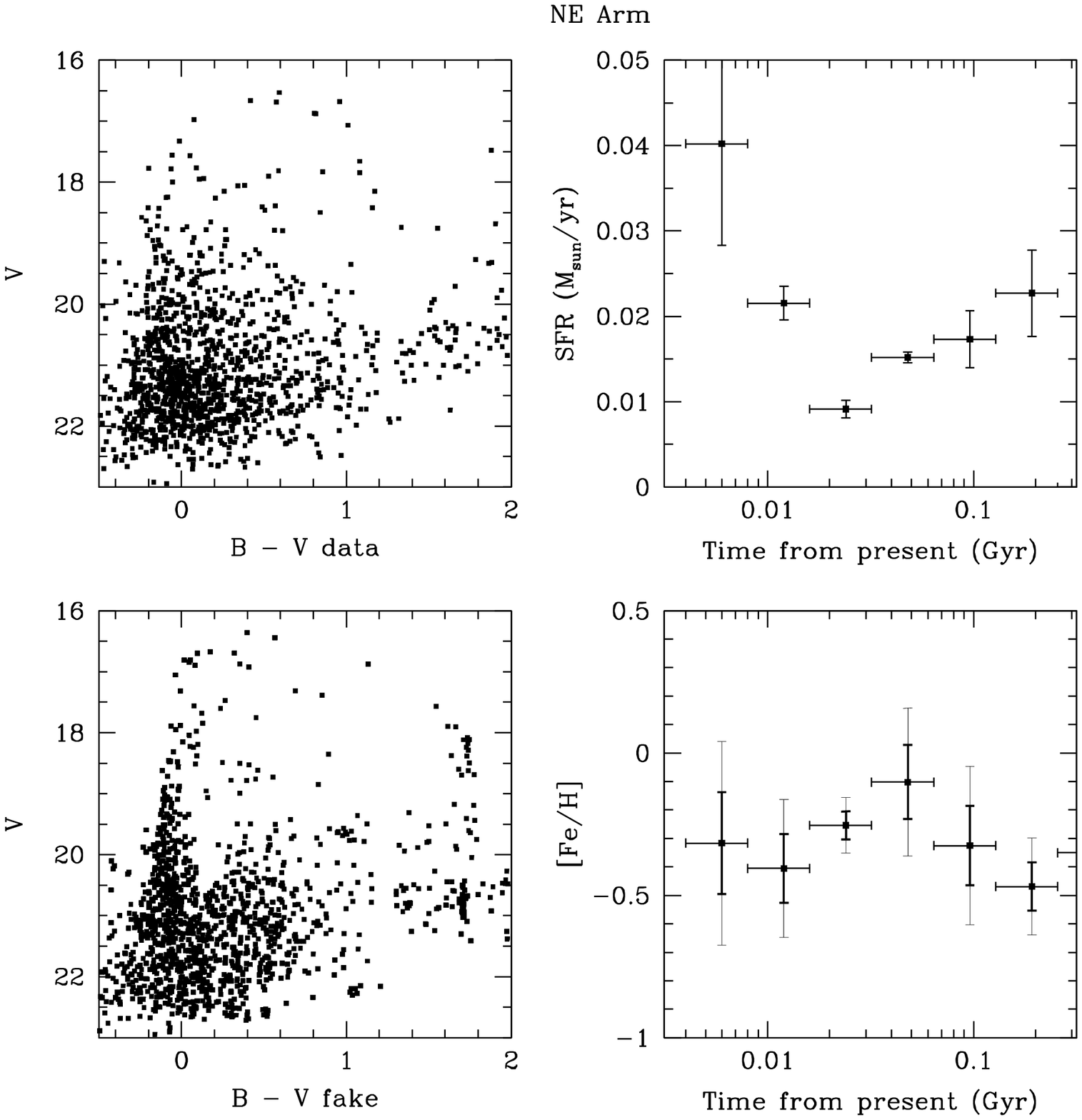,height=6.0in,angle=0}}
\caption{(b) The SFH of northeast spiral arm is shown.  The history reveals that the associations has been forming more stars than the surroundings for the past $\sim$100 Myr, as well as a very recent increase in the star formation rate.}
\end{figure}

\begin{figure}
\figurenum{11} 
\centerline{\psfig{file=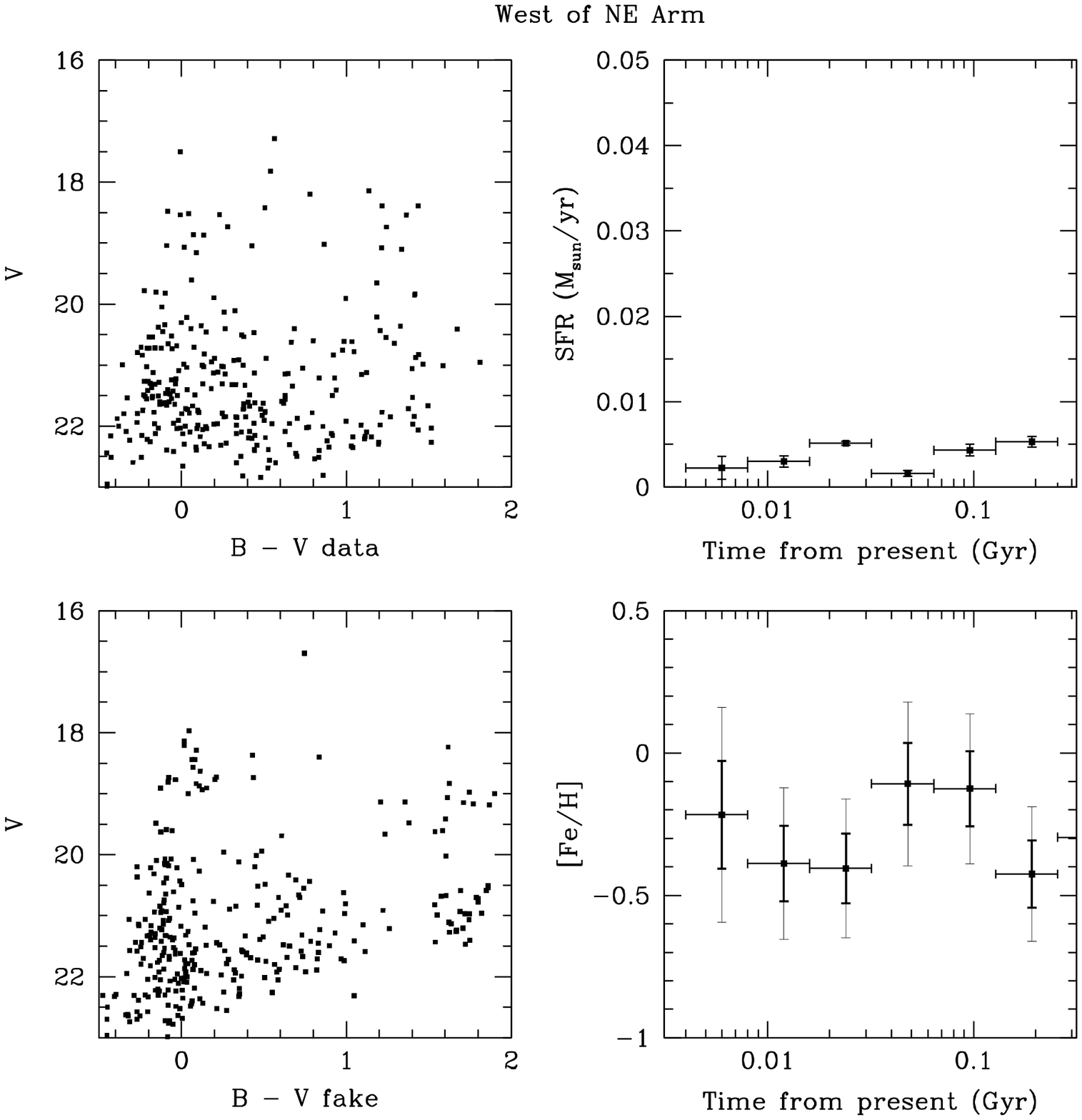,height=6.0in,angle=0}}
\caption{(c) The SFH of the region west of the arm is shown, and it similar to the region east of the association.}
\end{figure}

\end{document}